\documentclass[amsmath,amssymb,twocolumn,floatfix]{revtex4-2}

\usepackage{graphicx}
\usepackage{dcolumn}
\usepackage{bm}
\usepackage{hyperref}

\providecommand{\proarrow}[0]{\rightarrow}

\providecommand{\proname}[2]{#1 \proarrow #2}
\providecommand{\lrproname}[2]{#1 \leftrightarrow #2}

\providecommand{\gscat}[0]{\gamma_{\rm{scat}}}
\providecommand{\gabs}[0]{\gamma_{6}}
\providecommand{\cpa}[0]{\epsilon}
\providecommand{\cpmax}[0]{\epsilon^{\rm max}}

\begin{document}

\title{On low-scale baryogenesis from three-body decays}

\author{F.~Dom\'{\i}nguez}
\email{frandominguez@ugr.es}
\affiliation{Departamento de F\'{\i}sica At\'omica, Molecular y Nuclear, Universidad de Granada, 18071 Granada, Spain
}

\author{J.~Racker}
\email{jracker@unc.edu.ar}
\affiliation{Instituto de Astronom\'{\i}a Te\'orica y Experimental (IATE), Universidad Nacional
de C\'ordoba (UNC)~- Consejo Nacional de Investigaciones Cient\'{\i}ficas y T\'ecnicas
(CONICET), Laprida 854, X5000BGR, Córdoba, Argentina
}
\affiliation{
 Observatorio Astron\'omico de C\'ordoba (OAC), Universidad Nacional de C\'ordoba (UNC), Laprida 854, X5000BGR, Córdoba, Argentina
}

\begin{abstract}
Baryogenesis at the TeV scale from CP-violating decays of a massive particle requires some way to avoid the washouts from processes closely related to the existence of CP violation. It has been proposed that one way can be baryogenesis from three-body decays (instead of two-body decays). In this work we revisit this statement and show that, similarly to two-body-decay models, successful baryogenesis from three-body decays requires that the mass of the decaying particle be well above 10-100 TeV unless some other mechanism to avoid washouts is implemented. 
\end{abstract}

\maketitle

\section{Introduction}
\label{sec:introduction}
Models for explaining the baryon asymmetry of the universe via the CP-violating decay of a heavy particle typically involve high energy scales. This is because, on one hand, for certain models the connection with light neutrino masses imposes a lower bound for the mass $M$ of the decaying particle, like the Davidson-Ibarra bound~\cite{Davidson:2002qv} for type-I leptogenesis with hierarchical heavy neutrino masses, $M \gtrsim 10^8-10^9$~GeV (see also~\cite{Hambye:2003rt}). On the other hand, there is a more general issue arising from the washout processes related to the absorptive part of one-loop contributions to the CP asymmetry in decays. The strength of these processes is tied to the value of the CP asymmetry and their washout effect increases as the temperature during baryogenesis decreases (because the expansion rate of the universe becomes milder). It has been shown that this typically implies $M \gtrsim 10^5$~GeV for successful baryogenesis (see e.g.~\cite{Racker:2013lua}).

Masses as high as those given above might bring hierarchy problems~\cite{vissani97,clarke15}, be incompatible with cosmological scenarios that require low reheating temperatures and, more importantly, preclude experimental exploration in the foreseeable future. All this motivates research on baryogenesis models at or below the TeV scale (see e.g. the review~\cite{Chun:2017spz}), including ways to avoid the problems mentioned before for thermal baryogenesis from particle decays. For standard cosmological scenarios, with thermal baryogenesis occurring in a radiation dominated universe, three well-known ways (or mechanisms) have been identified and implemented in numerous models, namely:
(i) Enhancement of the CP asymmetry due to quasi-degenerate particles. This requires to have at least two particles with the same quantum numbers and very similar masses (according to the level of degeneracy oscillations may or may not be relevant in this case).
(ii)~Avoidance of washouts due to late decays, i.e. the particles decay and generate the asymmetry at temperatures well below $M$, when the rate of the washout processes has fallen to non-dangerous values. This case is not trivial to implement because the coupling responsible for the decay must be very small, therefore inverse decays and other production processes involving this coupling are suppressed, imposing the need for another interaction to produce the particles in the first place. Moreover, in order to allow for the late decays, this new interaction must not be active at temperature around or below $M$.
(iii) Boltzmann suppression of washouts when some of the decay products are massive (so that all relevant washout processes become Boltzmann suppressed). For baryogenesis above the electroweak phase transition 
this calls for the introduction of new fields (apart from the decaying particle that generates the asymmetry) and there is also another more subtle requirement discussed in~\cite{Racker:2013lua}.
A detailed joint discussion of these three mechanisms can be found in~\cite{Racker:2013lua}. We would also like to note that for very light decaying particles ($M$ around 100 GeV or below), (ii) and (iii) must be combined in order to avoid washouts (see the recent analysis of washouts in post-sphaleron baryogenesis in~\cite{Racker:2023yvv}).

In addition other ways have been proposed, particularly the one we are interested to analyze in this work, which is baryogenesis via three-body decays (instead of two-body decays), set forth in~\cite{hambye01} and implemented e.g.~in ~\cite{Gu:2011ff,Borah:2020ivi} (see~\cite{Fong:2013gaa} for another proposal and e.g.~\cite{PhysRevD.90.064050,*Dutta:2018zkg,*Chen:2019etb,*Mahanta:2019sfo,*Konar:2020vuu,*Chang:2021ose,*Chakraborty:2022gob,*DiMarco:2022doy,*Biswas:2023azl} for non-standard cosmological scenarios). Roughly the idea is that for three-body decays it is more easy to satisfy the out-of-equilibrium conditions and washouts are phase-space suppressed. The aim of our work is to show that this mechanism per se, i.e.~without also implementing some of the other ways mentioned above, does not allow to have baryogenesis at the TeV scale and, moreover, the bound $M \gtrsim 10^5$~GeV for successful baryogenesis also holds in this scenario. In order to analyze this mechanism we introduce a scalar model in Section~\ref{sec:model} to realize baryogenesis from three-body decays, then in Section~\ref{sec:bounds} we set the Boltzmann equations (BEs) in order to find a lower bound on $M$ for successful baryogenesis and finally we conclude in Section~\ref{sec:conclusions}.

\section{A model for baryogenesis from three-body decays}
\label{sec:model}
\begin{figure*}[t]
\includegraphics[width=0.33\textwidth]{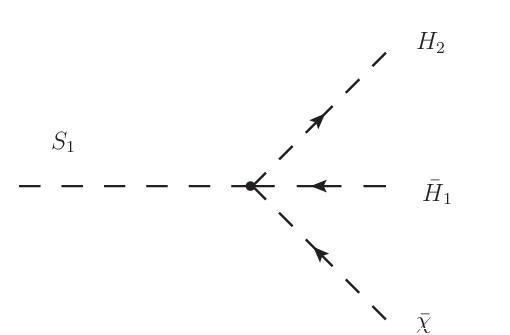} \quad
\includegraphics[width=0.46\textwidth,{angle=0}]{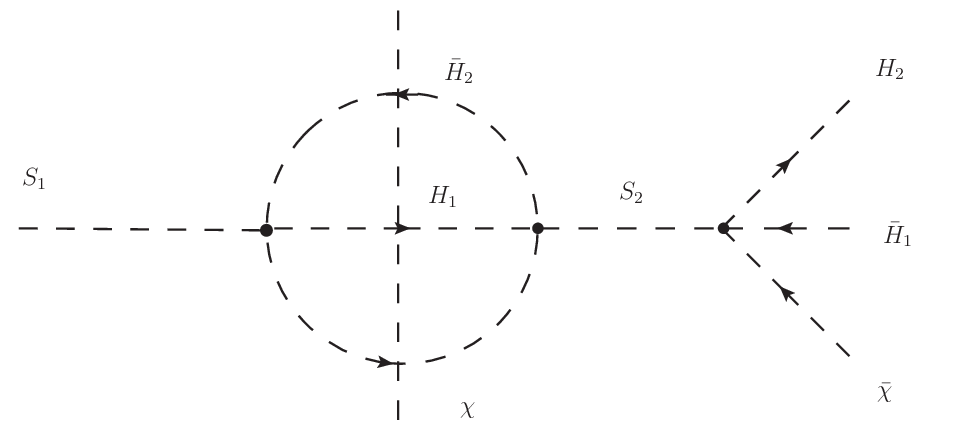}
\caption{\label{fig:cpasym} The interference of the tree-level diagram on the left and the one-loop diagram on the right can give rise to a CP asymmetry in the decays of $S_1$. The CP-odd phase arises when $\mathrm{Im}[(\lambda_1 \, \lambda_2^{\ast})^2] \ne 0$, 
while the CP-even phase when all the particles along the cut in the one-loop diagram can be on-shell. The process at the right of the cut may induce very efficient washouts at low temperatures and is responsible for the lower bound on $M_1$ discussed in this paper.}
\end{figure*}

The possibility to realize baryogenesis at the TeV scale from CP-violating three-body decays was suggested in~\cite{hambye01}. However, in the model proposed in~\cite{hambye01} as an illustration, all the operators in the Lagrangian that are relevant for baryogenesis involve three fields and the absorptive part of the one-loop diagrams contributing to the CP asymmetry involves $\proname{2}{2}$ processes which can washout the asymmetry very efficiently. For this class of three-body-decay models the analysis performed in~\cite{Racker:2013lua} can be applied because it involves the same kind of washout processes, with the conclusion that successful baryogenesis is not possible at scales much lower than $\sim$ 100~TeV (unless one of the mechanisms mentioned in the introduction is incorporated).
Therefore we proceed to build a model where all processes relevant for baryogenesis involve more particles than in standard baryogenesis from particle decays, and consequently they are phase-space suppressed. This might allow to satisfy more easily the out-of-equilibrium conditions and suppress washouts. In particular we want that only $\proname{3}{3}$ processes appear at the right of the cut in the one-loop diagrams, see Fig.~\ref{fig:cpasym}.

To this end we consider an extension of the Standard Model (SM) with a second Higgs doublet, $H_2$, two real scalar singlets, $S_i$ ($i=1,2$), and a complex scalar singlet field, $\chi$. 
In order to forbid two-body decays for the $S_i$, the model is supplemented with a discrete symmetry $Z_2$ and another exotic symmetry which can be a different discrete symmetry or a global $U(1)$. The charge assignments are given in Table~\ref{tab:charges}. We stress that our intention is not to propose a realistic model for baryogenesis, but to test the idea of three-body decays as a way for TeV scale baryogenesis, as explained above. Therefore we will not study issues like the vacuum stability of the scalar potential, breaking of the symmetries or the possibility to have a dark matter candidate (although the charges can be assigned in order to choose different possibilities for the lightest stable exotic particle(s)).
\begin{table}
\caption{\label{tab:charges}%
Charges of the fields under the exotic $Z_2$ and $U(1)$ symmetries. Depending on the choice of $Z_2$-charges for $H_2$ and $\chi$ it is possible to have different possibilities for the lightest stable exotic particle(s),
but we are not interested in analyzing this issue and any option is useful for our purposes.
Here ``SM'' represents all SM fields.
}
\begin{ruledtabular}
\begin{tabular}{lcc}
\textrm{ }&
\textrm{$Z_2$}&
\textrm{$U(1)$}\\
\colrule
SM & +1 & 0 \\
$S_{1,2}$ & -1&0 \\
$H_2$ & -1 (+1)& 1 \\
$\chi$ &+1 (-1)& 1\\
\end{tabular}
\end{ruledtabular}
\end{table}
The relevant terms for baryogenesis in the scalar potential $V$ are
\begin{eqnarray}
V(H_1,H_2,S_i,\chi) &=& \sum_i \frac{1}{2} M_i^2 S_i^2 + m_\chi^2 \chi^\dag \chi + m_{H_2}^2 H_2^\dag H_2 \nonumber \\ &+& \sum_i \lambda_i S_i H_2^\dag H_1 \chi + h.c. + \dots \,,
\end{eqnarray}
where $H_1$ is the SM Higgs.

Baryogenesis occurs through ``split Higgsogenesis''~\cite{davidson13}: an asymmetry initially develops in the scalar sector due to the CP-violating decays of the lightest real scalar singlet, $\proname{S_1}{H_2 \bar H_1 \bar \chi \, (\bar H_2 H_1 \chi)}$ and then it is partially transferred to baryons via fast Yukawa and sphaleron processes. 
In the following section we give a set of appropriate BEs and make a scan over the relevant parameters to determine the lowest value of $M_1$ that allows for successful baryogenesis without implementing any of the three mechanisms described in the introduction.
\section{Mass bounds for baryogenesis from three-body decays}
\label{sec:bounds}
Using Maxwell-Boltzmann statistics, assuming kinetic equilibrium, and working at linear order in the asymmetries, baryogenesis in the model described in the previous section can be described with the following set of transport equations:
\begin{eqnarray}
\frac{\mathrm{d} Y_{S_{1}}}{\mathrm{d}z}&=& - \frac{1}{zHs} \left(\frac{Y_{S_{1}}}{Y_{S_{1}}^{eq}}-1 \right)  \left( \gamma_{D} + \gscat \right) \: ,\nonumber \\
\frac{\mathrm{d} Y_{\Delta H_2}}{\mathrm{d}z} &=& \frac{1}{zHs} \left\{ \cpa \left(\frac{Y_{S_{1}}}{Y_{S_{1}}^{eq}}-1 \right)   \left( \gamma_{D} + \gscat \right)\right. \nonumber \\&& \left. - \left( y_{H_2} - y_{H_1} - y_{\chi} \right) \left[2\, \gabs + \frac{\gamma_{D}}{2} \right. \right. \nonumber \\&&  \left. \left. + \left(\frac{Y_{S_{1}}}{Y_{S_{1}}^{eq}}+2 \right) \frac{\gscat}{6} \right] \right\},
\label{eq:be}
\end{eqnarray}
where we have used the notation $H$ for the Hubble rate, $z\equiv M_1/T$ (with $T$ denoting the temperature), $Y_a \equiv \tfrac{n_a}{s}$ (with $n_a$ the number density of the particle specie ``$a$", and $s$ the entropy density), $Y_{\Delta a} \equiv Y_a - Y_{\bar a}$ (with $\bar{a}$ the CP-conjugate of $a$), and 
$y_a \equiv Y_{\Delta a}/Y_{a}^{eq}$ 
(with $Y_{a}^{eq}$ the equilibrium number density corresponding to one degree of freedom of particle ``$a$", normalized to the entropy density).

Since we want to determine the lowest value of $M_1$ compatible with successful baryogenesis without implementing any of the known mechanisms described in the introduction that allow for $M_1 \sim$ few TeV, i.e. without resorting to quasi-degenerate sates or massive decay products, we take $M_2 \gg M_1 \gg m_{H2}, m_\chi$. Then, at lowest non-zero order in $M_1/M_2$, the CP asymmetry in the decays of $S_1$, arising from the interference of tree-level and one-loop diagrams like the one depicted in Fig.~\ref{fig:cpasym}, is given by
\begin{eqnarray}
\cpa &\equiv & \frac{\gamma(S_1 \rightarrow H_2 \bar{H}_1 \bar{\chi})-\gamma(S_1 \rightarrow \bar{H}_2 H_1 \chi)}{\gamma(S_1 \rightarrow H_2 \bar{H}_1 \bar{\chi})+\gamma(S_1 \rightarrow \bar{H}_2 H_1 \chi)} \nonumber \\ &=& - \frac{3}{2^7 \pi^3} \, \frac{\mathrm{Im}[(\lambda_1 \, \lambda_2^{\ast})^2]}{|\lambda_1|^2} \, \frac{M_1^2}{M_2^2}\, .
\end{eqnarray}
When the relative phase of the couplings $\lambda_1$ and $\lambda_2$ is equal to $-\pi/4$, the CP asymmetry has a maximum value equal to
\begin{equation}
\cpmax = \frac{3}{2^7 \pi^3} \, |\lambda_2|^2 \, \frac{M_1^2}{M_2^2} = \frac{3}{2^7 \pi^3} \, \tilde{\lambda}^2,
\end{equation}
where we have defined the dimensionless parameter $\tilde{\lambda} \equiv \frac{|\lambda_2|}{M_2/M_1}$. 
Finally, for $M_2 \gg M_1 \gg m_{H2}, m_\chi$, the three reaction densities in these BEs are given at tree level by
\begin{eqnarray}
\gamma_{D}&=& \ \frac{1}{2^7 \pi^3} \ \frac{K_{1} (z)}{K_{2} (z)} \ n_{S_1}^{eq} \ \frac{M_1^2}{v^2} \  \tilde{m}, \\
\gscat &=& \frac{3}{(2 \pi)^5} \ \tilde{m} \ v^{-2} \  M_1^5 \ z^{-3} \ K_{1} (z),\\
\gabs &=& \frac{254}{(2 \pi)^9} \ \tilde{\lambda}^4 \  M_1^4 \ z^{-8}, 
\end{eqnarray}
whose magnitudes depend on two parameters that we have defined for convenience as
$\tilde{m} \equiv \frac{(|\lambda_1| \, v)^2}{M_1}$ (with $v$ denoting the Higgs vev, $v \simeq 174$ GeV) and $\tilde{\lambda}$ (introduced before). While $\gamma_D$ is just the reaction density of the decays of $S_1$, i.e. the total number of decays per unit time and volume (calculated at tree level and therefore equal to the inverse decays rate per unit volume), $\gscat$ and $\gabs$ are the sum of reaction densities of several processes with a similar role in baryogenesis and that consequently appear together in the BEs. In $\gscat $ we have summed the reaction densities of all $\proname{2}{2}$ scatterings involving $S_1$ in the initial state, i.e.~of $\proname{S_1 \chi}{\bar H_1 H_2}$ and all processes obtained by CP conjugation and/or interchanging initial and final states. Furthermore, $\gabs$ includes the reaction densities of all $\lrproname{3}{3}$ and $\lrproname{2}{4}$ processes mediated by $S_2$, i.e.~the one at the right of the cut in Fig.~\ref{fig:cpasym} and related processes obtained by CP conjugation and/or interchanging some of the initial and final states. All theses processes tend to washout the asymmetry and the tight connection between the magnitudes of $\cpmax$ and $\gabs$ is responsible for the lower bound on $M_1$ that we are going to calculate.

The BEs~\eqref{eq:be} can be complemented by a set of relations among chemical potentials and density asymmetries due to fast interactions and conserved charges, which lead to the equations
\begin{eqnarray}
y_{\chi} &=& -2 \, y_{H_2}, \nonumber \\
y_{H_1} &=& -\frac{13}{79}\, y_{H_2}, \nonumber \\
Y_B &=& \frac{6}{79}\, Y_{\Delta H_2},
\label{eq:asymrel}
\end{eqnarray}
with the first two equations allowing to solve the BEs for $Y_{\Delta H_2}$ and the last one giving the relation between the asymmetry in $H_2$ and the baryon asymmetry normalized to the entropy density (see~\cite{davidson13}).

\begin{figure}[t]
\includegraphics[width=0.33\textwidth,{angle=270}]{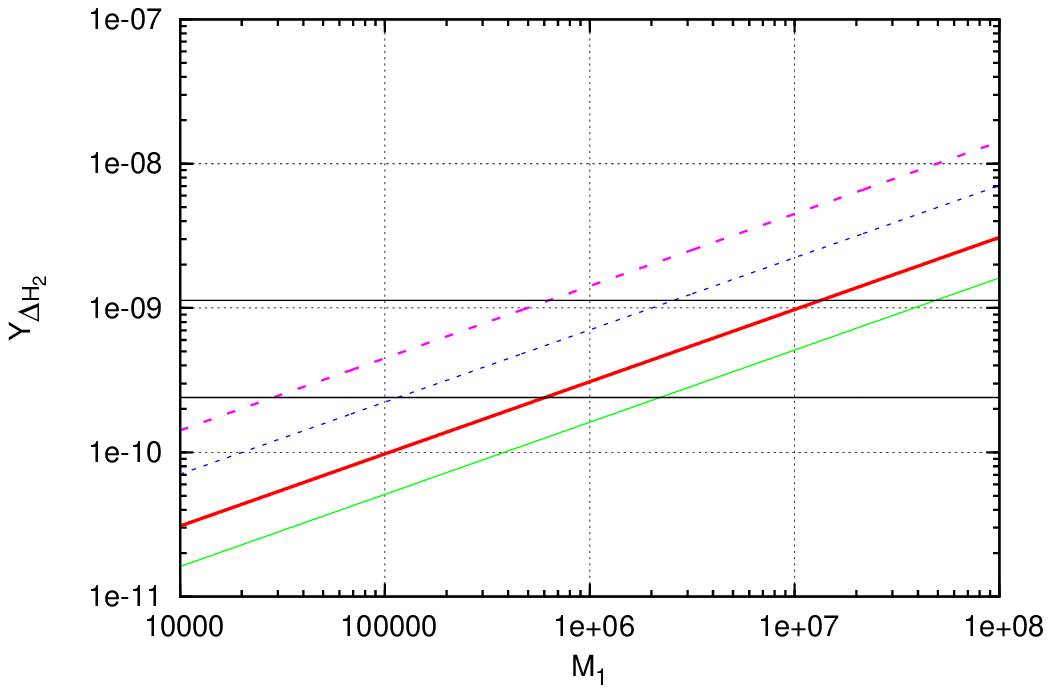}
\caption{\label{fig:ybmax} Maximum $H_2$ asymmetry as a function of $M_1$ (in GeV). 
The solid thick red line gives the asymmetry for the model considered in this paper, the solid thin green line corresponds to the same model but without including the $\proname{2}{2}$ scatterings, i.e.~setting $\gscat = 0$ in the BEs (see discussion in the text), and the pink-dashed (blue-dotted) line gives the asymmetry for a toy scenario where there is only one process and channel contributing to the CP asymmetry and the washouts mediated by $S_2$, including (not including) the $\proname{2}{2}$ scatterings.
The upper horizontal solid black line indicates the value $Y_{\Delta H_2}$ must have to obtain the observed baryon asymmetry in our model, while the lower horizontal one the value it should have were the conversion factor between $Y_B$ and $Y_{\Delta H_2}$ equal to 28/79 instead of 6/79 (drawn to ease comparisons with more standard leptogenesis models).}
\end{figure}

Next we proceed to determine the maximum amount of asymmetry that can be generated in the $H_2$ field as a function of $M_1$, i.e. we maximize the final value of $Y_{\Delta H_2}$ over the two free parameters $\tilde \lambda$ and $\tilde m$ for each value of $M_1$. In order to achieve this, we integrate the BEs~\eqref{eq:be} taking $\cpa=\cpmax$ and $Y_{\Delta H_2} (z \ll 1) = Y_{S_1}(z \ll 1) = 0$ as initial conditions. Note that we take an initial zero abundance for $S_1$ because we do not want to study the lower bound on $M_1$ in scenarios where the $S_1$ could have been produced by other (CP-conserving) processes, since in this case it is known that $M_1$ can be in the TeV scale (via the mechanism (ii) described in the introduction). The result is represented by the thick solid red line in Fig.~\ref{fig:ybmax}.
Taking into account the relation between $Y_{\Delta H_2}$ and $Y_B$ given in Eq.~\eqref{eq:asymrel}, we conclude that the lower bound on $M_1$ for successful baryogenesis in this model is very high, $M_1 \gtrsim 10^7$~GeV. This is actually much higher than the lower bound  $M \sim 10^5$~GeV found in~\cite{Racker:2013lua} for two-body decays (without the implementation of the mechanisms (i)-(iii) outlined in the introduction). Also note that the maximum value of $Y_{\Delta H_2}$ grows as the square root of $M_1$ (see~\cite{Racker:2013lua} for a more detailed discussion on parameter dependences in two-body-decay models).

The reason for the higher bound on $M_1$ compared to the two-body-decay scenario is twofold. On one hand, in the particular model we have chosen, the asymmetry generated in decays originates in the scalar sector, particularly in $H_2$, and is partially transferred and shared among many fields. Therefore the conversion factor between $Y_{\Delta H_2}$ and $Y_B$ is significantly smaller (around four to five times smaller) than the corresponding conversion factor in more standard baryogenesis-via-leptogenesis models. However, even if this conversion factor were as big as in leptogenesis models ($\sim 1/3$), the lower bound on $M_1$ would still be well above $10^5$~GeV (the two horizontal solid black lines in Fig.~\ref{fig:ybmax} correspond to the values $Y_{\Delta H_2}$ must have so that the baryon asymmetry equals the observed value, $Y_B^{\rm obs} \simeq 8.6 \times 10^{-11}$, for conversion factors equal to  $6/79$ and $28/79$). On the other hand, the four-field operators allow for many $\lrproname{3}{3}$ and $\lrproname{2}{4}$ washout processes, and with several channels contributing to each of them, resulting in an enhancement of the washout rate relative to the CP asymmetry when compared to two-body-decay scenarios. In order to quantify this effect, we have also plotted in Fig.~\ref{fig:ybmax} the maximum value of $Y_{\Delta H_2}$ that would be obtained if only one process and channel contributed to the CP asymmetry and the related washout rate $\gabs$ (see the pink dashed line). Again it can be seen that, even in this toy scenario, the lower bound on $M_1$ is well above $10^5$~GeV (or $10^4$~GeV for a larger conversion factor), reinforcing the conclusion that the three-body-decay mechanism per se does not allow for baryogenesis with $M_1$ at the TeV scale.  

It is interesting to note that in the model we are analyzing, the same operator in the Lagrangian that is responsible for the three-body decays of $S_1$, allows for $\proname{2}{2}$ scatterings processes, which are much more efficient than inverse decays ($\proname{H_2 \bar H_1 \bar \chi, \, \bar H_2 H \chi}{S_1}$) to produce the $S_1$ at high temperatures (while becoming subdominant at temperatures somewhat below $M_1$). 
This may allow to choose smaller values of $\lambda_1$ to delay the decays (reducing washout effects on the asymmetry), without compromising too much the production of $S_1$, and consequently realizing a late decay scenario without the need for an extra interaction to produce $S_1$ at high temperatures (see (ii) in the introduction). Indeed this is a relevant effect, although, as we have stated above, the bound on $M_1$ stays well above $10^4-10^5$~GeV. To demonstrate this point we have drawn the solid thin green line in Fig.~\ref{fig:ybmax}, which gives the maximum value of $Y_{\Delta H_2}$ without including the $\proname{2}{2}$ scatterings processes in the BEs (i.e. taking $\gscat=0$). It can be seen that if it were not for this effect the lower bound on $M_1$ would be a factor 3 to 4 larger than quoted previously (for comparisons we have also depicted with the blue dotted line the maximum value of $Y_{\Delta H_2}$ taking $\gscat=0$ in the toy scenario with only one process and channel contributing to the CP asymmetry and the related washout rate). 

Before concluding we wish to make one more remark. For simplicity (considering in particular the large number of processes involved), we have worked in the hierarchical limit $M_2 \gg M_1$. According to the analysis of~\cite{Racker:2013lua} for a two-body-decay model, the bound on $M_1$ could be somewhat lower for $M_2 \sim \text{5--10} \, M_1$. However the difference with respect to the hierarchical limit is mild (less than a factor 2 for the inert doublet model studied in~\cite{Racker:2013lua}), which does not alter the conclusion of this work (moreover, note that to find the bound when $M_2$ is closer to $M_1$ would require to include the evolution of $Y_{S_2}$ in the BEs, together with processes with $S_2$ on-shell).

\section{Conclusions}
\label{sec:conclusions}

Baryogenesis from particle decays can be realized in a host of models beyond the SM of particle physics. For the reasons explained in the introduction, typically the mass $M$ of the decaying particle must be very high, $M \gtrsim 10^5$~GeV, or even much larger. However, there are theoretical and experimental motivations to explore models with lower masses, at the TeV scale or below. In order to avoid the washout of the asymmetry from processes closely related to the existence of CP violation in decays, three ways or mechanisms have been implemented in many models of baryogenesis from particle decays at or below the TeV scale (enumerated as (i) to (iii) in the introduction). In addition there have been other proposals, one of which has been the focus of this work, namely baryogenesis from three-body decays (instead of two-body decays).

When the washout processes closely related to the CP asymmetry (i.e.~the ones at the right of the cut in one-loop diagrams) are $\proname{2}{2}$ scatterings, it seems clear that the analysis in~\cite{Racker:2013lua} can be applied and a lower bound on $M$ around $10^5$~GeV must hold. Therefore we have proposed another model in Section~\ref{sec:model} to explore the possibility of achieving baryogenesis at the TeV scale from three-body decays and where all relevant processes for baryogenesis, particularly the most dangerous washout processes, are phase-space suppressed compared to more standard baryogenesis models. Our results show that the mechanism of baryogenesis from three-body decays in itself does not allow to circumvent the lower bound $M \gtrsim 10^5$~GeV obtained e.g.~in~\cite{Racker:2013lua} (see Fig.~\ref{fig:ybmax}). This does not imply that baryogenesis from three-body decays is not possible at the TeV scale, but that the problem with washout processes is akin to two-body-decay scenarios and that one of the mechanisms (i), (ii) or (iii) should be implemented for masses below $\sim 10^4-10^5$~GeV. Given the present lack of experimental data to probe many models of baryogenesis, the theoretical constrain for a class of models that we have found might actually be seen as a positive result.

\begin{acknowledgments}
F. D. acknowledges support from the Spanish MICINN under Contract No. PTA2018-016573-I. J. R. thanks Juan Herrero-Garc\'{\i}a for useful comments.
\end{acknowledgments}

\bibliography{3bodybaryo}

\end{document}